\documentclass[twocolumn,aps,superscriptaddress,showpacs,nofootinbib,floatfix]{revtex4}
\usepackage{epsfig,bm,feynmf}
\usepackage{graphics}
\usepackage{amsmath}
\usepackage[normalem]{ulem}  
\usepackage[dvips]{color} 

\renewcommand\sout{\bgroup \color{red} \ULdepth=-.5ex \ULset}
\begin{document}


\title{Low-energy Bremsstrahlung photon in relativistic nucleon+nucleon collisions}


\author{Taesoo Song}\email{taesoo.song@theo.physik.uni-giessen.de}
\affiliation{Institut f\"{u}r Theoretische Physik, Universit\"{a}t
Gie\ss en, Germany}


\author{Pierre Moreau}
\affiliation{Institute for Theoretical Physics, Johann Wolfgang
Goethe Universit\"{a}t, Frankfurt am Main, Germany}



\begin{abstract}
We study the production of Bremsstrahlung photon in relativistic nucleon+nucleon collisions by introducing a deceleration time of electromagnetic currents.
It is found that Bremsstrahlung photon spectrum at low energy does not depend on the deceleration time but solely on the amount of reduced electromagnetic current in collision.
On the other hand, the photon spectrum becomes soft with increasing deceleration time.
We also find that Bremsstrahlung photon spectrum in p+n collisions is considerably different from that in p+p collisions at low energy.
\end{abstract}


\maketitle

\section{Introduction}

Relativistic heavy-ion collisions are presently the unique way to produce an extremely hot and dense nuclear matter in laboratory and to study the properties of such a matter.
There are several kinds of probe particles through which one can investigate the properties and electromagnetic probe is one of them.
The electromagnetic probe is distinguished from other particles in a couple of respects.
First of all, it has no color charge and interacts only through electromagnetic coupling which is much weaker than strong coupling, that it gets out of the nuclear matter without any further interactions, after it is produced.
Secondly, it is continuously produced from initial hard scattering to after freeze-out in relativistic heavy-ion collisions.

Photon in relativistic heavy-ion collisions is classified into three parts according to production stage.
The first part of photon is produced through initial hard scattering which can be obtained by rescaling photon spectrum in nucleon+nucleon collisions with the number of binary collisions.
Then produced nuclear matter emits thermal photon both in QGP and in hadron gas phase.
The third part of photon is produced through the electromagnetic decay of hadrons mostly after the freeze-out.
The first two parts of photon are called direct photon and the last one indirect photon.

Nucleon+nucleon collision are reference experiment to study the nuclear matter produced in heavy-ion collisions, because they are hardly able to produce a sizable matter, unless the collision happens near the Large Hadron Collider (LHC) energies.
For example, direct photons in p+p collision subtracted from those in heavy-ion collision with the number of binary collisions multiplied, the leftovers are interpreted as thermal photons~\cite{Adare:2014fwh}.
In this respect photon production in nucleon+nucleon collisions is the first step to study the nuclear matter produced in relativistic heavy-ion collisions.

In microscopic view direct photon is produced in nucleon+nucleon collisions through the scattering of quarks and antiquarks composing the nucleons, for example, $q+\bar{q}\rightarrow  g+\gamma$, $q(\bar{q})+g\rightarrow  q(\bar{q})+\gamma$, and so on.
These elementary scattering cross sections are convoluted with parton distribution functions of nucleon.
However, this perturbative Quantum Chromodynamics (pQCD) approach with the factorization formula is reliable only for large energy momentum transfer, in other words, for the production of high-energy photon.

Another source of the direct photon is the Bremsstrahlung, which is the electromagnetic radiation from decelerated charged particles.
The `Brems' means `to brake' from the German word `bremsen' and `strahlung' means radiation.
Microscopically the Bremsstrahlung photon is produced through the parton scatterings, $q+\bar{q}\rightarrow q+\bar{q}+\gamma$ or $q+q(g)\rightarrow q+q(g)+\gamma$ or $\bar{q}+\bar{q}(g)\rightarrow \bar{q}+\bar{q}(g)+\gamma$.
However, if the energy of emitted photon is low, it looks like $N+N \rightarrow N+N+\gamma$ in low-energy collisions or $N+N \rightarrow X+X^\prime+\gamma$ in high-energy collisions, where $X$ and $X^\prime$ represent wounded nucleons which still carry electromagnetic currents in beam direction.

If the collision energy is extremely large, two nucleons pass through each other, and the stopping or deceleration of electromagnetic currents can be reduced to one-dimensional problem.
Since nucleon is not a point-like particle but a composite particle and abundant particles are produced in high-energy collisions, the stopping should be described by a smooth function of time.
In this study we model the production of low-energy Bremsstrahlung photon in relativistic nucleon+nucleon collisions, introducing a finite stopping time of electromagnetic currents.

We first describe in section~\ref{stopping} the stopping of charged particles from the simplest case to more sophisticated ones step by step, and take into account the structure of nucleon in section~\ref{structure}. After that the stopping or deceleration of charged particle takes place by collisions in section~\ref{collision}, and the results are applied to relativistic nucleon+nucleon collisions in section~\ref{experiment}. Finally summary is given in section~\ref{summary} and several useful Fourier transformations are presented in the Appendix.

\section{Stopping of charged particle}
\label{stopping}

\subsection{no stopping}
\label{nostop}

The momentum distribution of radiated photon from decelerated charged particle is expressed as~\cite{Koch:1990jzd,Itzykson:1980,Biro:2014dja}
\begin{eqnarray}
\omega\frac{dN}{d^3 {\bf k}}=\frac{1}{2(2\pi)^3}\sum_\lambda|j(\omega,{\bf k})\cdot\epsilon^\lambda(\omega,{\bf k})|^2,
\label{bremsstrahlung}
\end{eqnarray}
where $\omega,~{\bf k}$ are photon energy and momentum, $j(\omega,{\bf k})$ electromagnetic current, and $\epsilon^\lambda(\omega,{\bf k})$ the polarization vector of emitted photon with $\lambda$ being polarization state.

As a warm-up we consider a particle with electric charge $Q$ and velocity $v$ without stopping or deceleration.
The electromagnetic current is then given by
\begin{eqnarray}
{\bf j}(t,{\bf r})=Qv\delta(z-vt)\delta(x)\delta(y)~{\bf e_z},
\label{current0}
\end{eqnarray}
where ${\bf e_z}$ is the unit vector in z-direction.
Position of the charged particle Fourier-transformed,
\begin{eqnarray}
{\bf j}(t,{\bf k})=\int d^3{\bf r} {\bf j}(t,{\bf r})e^{i {\bf r}\cdot{\bf k}}=Qv e^{ik_z vt}~{\bf e_z},
\label{fourier-r}
\end{eqnarray}
and $t$ transformed into $\omega$,
\begin{eqnarray}
\int dt {\bf j}(t,{\bf k})e^{-i\omega t}=Qv \int dt e^{i(k_z v-\omega)t}~{\bf e_z}\nonumber\\
=2\pi Qv\delta(k_z v-\omega)~{\bf e_z}.
\label{fourier-t}
\end{eqnarray}

Substituting Eq.~(\ref{fourier-t}) into Eq.~(\ref{bremsstrahlung}),
the spectrum of radiated photon is given by
\begin{eqnarray}
\omega\frac{dN}{d^3{\bf k}}=\frac{1}{4\pi}\sum_\lambda\bigg\{Qv\epsilon_z^\lambda \delta(k_z v-\omega)\bigg\}^2\nonumber\\
=\frac{1}{4\pi}\sum_\lambda\bigg\{Qv\epsilon_z^\lambda \delta[\omega( v\cos\theta-1)]\bigg\}^2
\end{eqnarray}
where we use the Coulomb gauge ($\epsilon^0=0$) and $\theta$ is the angle of ${\bf k}$ with respect to ${\bf e_z}$.
Since charged particle in nature always has a nonvanishing mass, it cannot reach the speed of light ($v<1)$. Therefore,
\begin{eqnarray}
\omega\frac{dN}{d^3{\bf k}}=0.
\end{eqnarray}

Now we deal with the stopping of a charged particle from the simplest case to sophisticated ones step by step.

\subsection{instant stopping}

In the simplest case a particle with the constant velocity $v$ and the electric charge $Q$ instantly stops at $(t,{\bf r})=(0,0)$.
The current is described by
\begin{eqnarray}
{\bf j}(t,{\bf r})=Qv\delta(z-vt)\delta(x)\delta(y)\theta(-t)~{\bf e_z},
\label{instantjr}
\end{eqnarray}
which is same as Eq.~(\ref{current0}) except the step function $\theta(-t)$.
Carrying out Fourier-transformations,
\begin{eqnarray}
{\bf j}(t,{\bf k})=\int d^3{\bf r} {\bf j}(t,{\bf r})e^{i{\bf r\cdot k}}=Qv \theta(-t)e^{ik_z vt}~{\bf e_z},
\end{eqnarray}
and

\begin{eqnarray}
\int dt {\bf j}(t,{\bf k})e^{-i\omega t}=Qv \int dt \theta(-t)e^{i(k_z v-\omega)t}~{\bf e_z}\nonumber\\
=Qv\bigg\{ \pi\delta(\omega-k_zv)+\frac{i}{\omega-k_zv}\bigg\}~{\bf e_z},
\label{instantjk}
\end{eqnarray}
by using Eq.~(\ref{stepf}) and the relation
\begin{eqnarray}
\theta(-t)=\frac{1}{2}\{1-{\rm sgn}(t)\},
\end{eqnarray}
where ${\rm sgn}(t)$ is signum function.
Dropping off the delta function in Eq.~(\ref{instantjk}), the spectrum of emitted photon is given by
\begin{eqnarray}
\omega\frac{dN}{d^3{\bf k}}=\frac{1}{2(2\pi)^3}\sum_\lambda\bigg(\frac{Qv\epsilon_z^\lambda}{\omega-k_zv}\bigg)^2,
\label{instantg}
\end{eqnarray}
which is in covariant form~\cite{Haglin:1992fy,Eggers:1995jq}

\begin{eqnarray}
\omega\frac{dN}{d^3{\bf k}}=\frac{1}{2(2\pi)^3}\sum_\lambda Q^2\bigg(\frac{p\cdot \epsilon^\lambda}{p\cdot k}\bigg)^2,
\end{eqnarray}
where $p$ is the four momentum of charged particle.

Now we turn to the polarization vector of photon.
Suppose photon momentum and a polarization vector are respectively expressed as
\begin{eqnarray}
\vec{k}=\begin{pmatrix} 0, & 0, & k \end{pmatrix},\nonumber\\
\vec{\epsilon}^1=\begin{pmatrix} \cos\varphi, & \sin\varphi, & 0 \end{pmatrix},
\end{eqnarray}
where $\varphi$ is the polarization angle.
Rotating $\vec{k}$ by an angle $\theta$ around y-axis and then by an angle $\phi$ around z-axis,

\begin{eqnarray}
\vec{k}^\prime=
\begin{pmatrix} \cos\phi & -\sin\phi & 0 \\ \sin\phi & \cos\phi & 0 \\ 0 & 0 & 1 \end{pmatrix}
\begin{pmatrix} \cos\theta & 0 & \sin\theta \\ 0 & 1 & 0 \\ -\sin\theta & 0 & \cos\theta \end{pmatrix}
\begin{pmatrix} 0 \\ 0 \\ k \end{pmatrix}\nonumber\\
=\begin{pmatrix} k\sin\theta\cos\phi \\ k\sin\theta\sin\phi \\ k\cos\theta \end{pmatrix},~~~\nonumber\\
\vec{\epsilon}^{~1\prime}=
\begin{pmatrix} \cos\phi & -\sin\phi & 0 \\ \sin\phi & \cos\phi & 0 \\ 0 & 0 & 1 \end{pmatrix}
\begin{pmatrix} \cos\theta & 0 & \sin\theta \\ 0 & 1 & 0 \\ -\sin\theta & 0 & \cos\theta \end{pmatrix}
\begin{pmatrix} \cos\varphi \\ \sin\varphi \\ 0 \end{pmatrix}\nonumber\\
=\begin{pmatrix} \cos\theta\cos\phi\cos\varphi-\sin\phi\sin\varphi \\ \cos\theta\sin\phi\cos\varphi+\cos\phi\sin\varphi \\ -\sin\theta\cos\varphi \end{pmatrix}.~~~
\end{eqnarray}

Since $\epsilon^1\cdot {\bf e_z}=-\sin\theta\cos\varphi$, the average of $(\epsilon^1\cdot {\bf e_z})^2$ in Eq.~(\ref{instantg}) over $\varphi$  turns to

\begin{eqnarray}
\frac{1}{2\pi}\int_0^{2\pi} d\varphi~(\epsilon^1\cdot {\bf e_z})^2= \frac{1}{2}\sin^2\theta,
\end{eqnarray}
and
\begin{eqnarray}
\frac{1}{2\pi}\sum_{\lambda=1,2}\int_0^{2\pi} d\varphi~(\epsilon^\lambda\cdot {\bf e_z})^2= \sin^2\theta,
\end{eqnarray}
for $\epsilon_2$ has the same contribution as $\epsilon_1$.
The same result is obtained by using the relation

\begin{eqnarray}
\sum_{\lambda=1,2}\epsilon_i^\lambda \epsilon_j^{\lambda*}=\delta_{ij}-\hat{k}_i\hat{k}_j,
\end{eqnarray}
where $\hat{k}_i=k_i/|{\bf k}|$.
Therefore, at mid-rapidity ($\sin\theta=1$)

\begin{eqnarray}
\frac{dN}{d^2k_T dy}\bigg|_{y=0}=\frac{1}{2(2\pi)^3}\bigg(\frac{Qv}{\omega}\bigg)^2,
\label{instantg2}
\end{eqnarray}

where $\sum_\lambda(\epsilon^\lambda\cdot {\bf e_z})^2$ is substituted with $1$, and it will be applied throughout this paper.

\subsection{smooth stopping}
Next we deal with smooth stopping by using hyperbolic tangent function instead of step function in Eq.~(\ref{instantjr}):

\begin{eqnarray}
{\bf j}(t,{\bf r})=Qv\delta(z-vt)\delta(x)\delta(y)\frac{1-\tanh(at)}{2}~{\bf e_z},
\end{eqnarray}
where for simplicity we assumed that the particle keeps its initial velocity but the electric charge evaporates with time and $a$ controls evaporation time.

The electromagnetic current is modified by Fourier transformations into

\begin{eqnarray}
{\bf j}(t,{\bf k})=Qv \frac{1-\tanh(at)}{2}e^{ik_z vt}~{\bf e_z},
\end{eqnarray}
and

\begin{eqnarray}
\int dt {\bf j}(t,{\bf k})e^{-i\omega t}=Qv \int dt \frac{1-\tanh(at)}{2}e^{i(k_z v-\omega)t}~{\bf e_z}\nonumber\\
=Qv\bigg\{ \pi\delta(\omega-k_zv)+\frac{\pi i}{2a}{\rm csch}\bigg(\frac{\pi (\omega-k_zv)}{2a}\bigg)\bigg\}~{\bf e_z}~~
\label{smoothj}
\end{eqnarray}
by using Eq.~(\ref{tanh}).
Ignoring the unphysical pole at $\omega=k_zv$,

\begin{eqnarray}
\omega\frac{dN}{d^3k}\bigg|_{y=0}=\frac{1}{2(2\pi)^3}\bigg(\frac{\pi Qv}{2a}\bigg)^2{\rm csch}^2\bigg(\frac{\pi \omega}{2a}\bigg).
\label{smoothg}
\end{eqnarray}

We can find that Eq.~(\ref{smoothg}) converges into Eq.~(\ref{instantg2}) in the limit of $a\rightarrow \infty$, because

\begin{eqnarray}
\lim_{x\rightarrow 0}{\rm csch}~x=\frac{1}{x}.
\end{eqnarray}

\begin{figure} [h]
\centerline{
\includegraphics[width=8.6 cm]{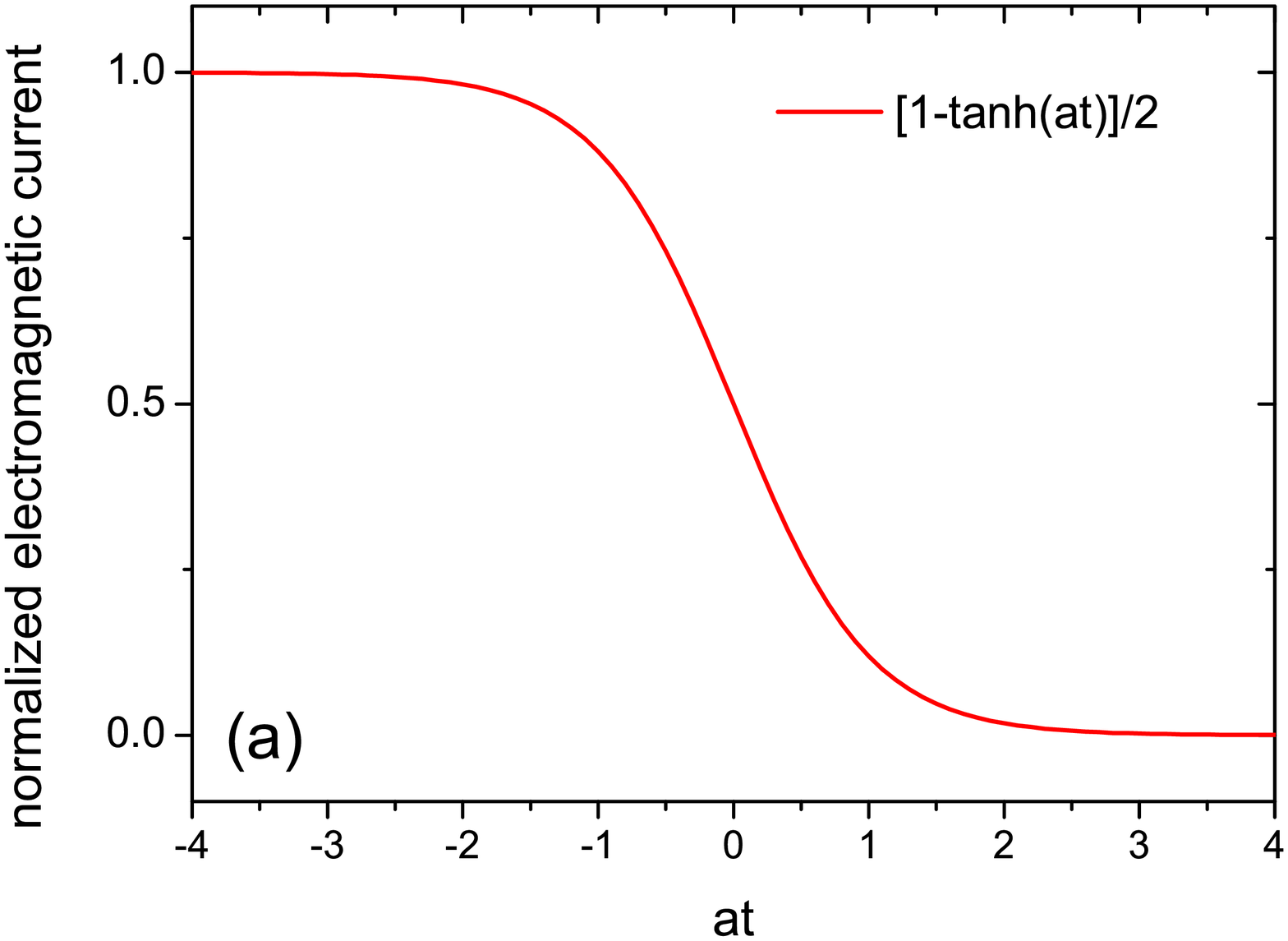}}
\centerline{
\includegraphics[width=8.6 cm]{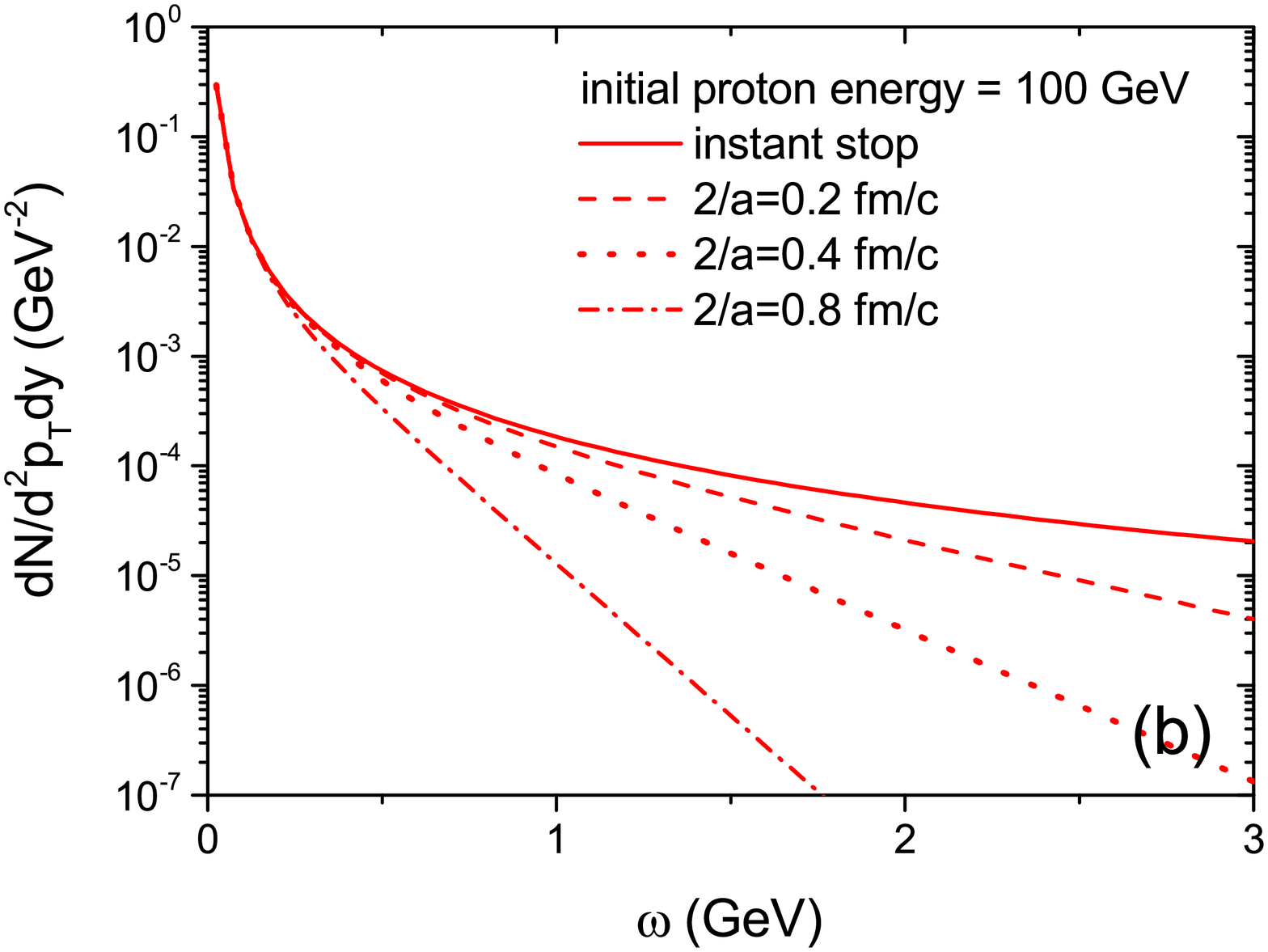}}
\caption{(Color online) (a) normalized electromagnetic current as a function of rescaled time and (b) photon spectra at mid-rapidity for various stopping times of the proton whose initial energy is 100 GeV} \label{smooth1}
\end{figure}

The upper panel of figure~\ref{smooth1} shows normalized electromagnetic current as a function of rescaled time.
Since we use hyperbolic tangent function, the deceleration of charged particle starts very early.
However, we can define the effective stopping time as $2/a$ during which electromagnetic current decreases from 88 \% to 12 \%.

We show in the lower panel photon spectra at mid-rapidity for various stopping times of a proton whose initial energy is 100 GeV.
$\alpha=e^2/(4\pi)$ is taken to be 1/137.
Though proton is not a point-like particle, its detailed structure is ignored for simplicity.
The figure clearly shows that as the stopping time increases, the photon spectrum becomes soft.
One interesting point is that the photon spectrum at very low frequency does not depend on the stopping time.
We will explain the reason for it in the following subsection.

\subsection{stepwise stopping}

Now we describe the smooth stopping by discretizing the process.
As the first trial, we assume a charged particle changes its velocity from $v_i$ to $v_f$ at $(t,z)=(0,0)$. Then electromagnetic current is expressed as
\begin{eqnarray}
{\bf j}(t,{\bf r})=Qv(t)\delta(z-v(t)t)\delta(x)\delta(y)~{\bf e_z},
\end{eqnarray}
where $v(t)=v_i\theta(-t)+v_f\theta(t)$.
After Fourier transformations the current turns to

\begin{eqnarray}
{\bf j}(t,{\bf k})=\int d^3{\bf r} {\bf j}(t,{\bf r})e^{i{\bf r\cdot k}}=Qv(t) e^{ik_z v(t)t}~{\bf e_z},
\end{eqnarray}
and

\begin{eqnarray}
\int dt {\bf j}(t,{\bf k})e^{-i\omega t}=Q\bigg\{v_i \int dt \theta(-t) e^{i(k_z v_i-\omega)t}\nonumber\\
+v_f \int dt \theta(t) e^{i(k_z v_f-\omega)t}\bigg\}~{\bf e_z}\nonumber\\
=Q\bigg\{\frac{iv_i}{\omega-k_zv_i}-\frac{iv_f}{\omega-k_zv_f}\bigg\}~{\bf e_z},
\label{stepj1}
\end{eqnarray}
where we removed insignificant delta functions and will do so from here on.
We can find that Eq.~(\ref{stepj1}) is equivalent to Eq.~(\ref{instantjk}) in case of $v_f= 0$.

Now we suppose the charged particle is decelerated in two steps, that is, the velocity changes from $v_i$ to $v_m$ at $t=-T/2$ and then to $v_f$ at $t=T/2$:

\begin{eqnarray}
v(t)=v_i\theta(-t-T/2)+v_m{\rm rect}_T(t) +v_f\theta(t-T/2),
\end{eqnarray}
with ${\rm rect}_T(t)$ being the box function defined in Eq.~(\ref{boxf}), and the position of charged particle,
\begin{eqnarray}
z(t)=\int dt v(t)=v_m t ~~{\rm for}~ |t|\leq T/2,\nonumber\\
=v_m T/2 +v_f(t-T/2) ~~{\rm for}~ t> T/2,\nonumber\\
=-v_m T/2 +v_i(t+T/2) ~~{\rm for}~ t< -T/2.
\end{eqnarray}

Carrying out Fourier transformations,
\begin{eqnarray}
{\bf j}(t,{\bf k})=\int d^3{\bf r} {\bf j}(t,{\bf r})e^{i{\bf r\cdot k}}=Qv(t) e^{ik_z z(t)}~{\bf e_z},
\end{eqnarray}
and
\begin{eqnarray}
&&\int dt {\bf j}(t,{\bf k})e^{-i\omega t}\nonumber\\
&&=Q v_i e^{i(\omega-k_z v_m)T/2}\int dt' \theta(-t')e^{i(k_z v_i-\omega)t'}~{\bf e_z}\nonumber\\
&&+Q v_m \int dt ~{\rm rect}_T(t)e^{i(k_z v_m-\omega)t}~{\bf e_z}\nonumber\\
&&+Q v_f e^{i(k_z v_m-\omega)T/2}\int dt'' \theta(t'')e^{i(k_z v_f-\omega)t''}~{\bf e_z},
\end{eqnarray}
where $t'=t+T/2$ and $t''=t-T/2$.
By using Eq.~(\ref{stepf}),

\begin{eqnarray}
{\bf j}(\omega,{\bf k})=iQ\bigg[ \bigg(\frac{v_i}{\omega-k_z v_i} - \frac{v_m}{\omega-k_z v_m}\bigg)e^{i(\omega-k_z v_m)T/2}\nonumber\\
+\bigg(\frac{v_m}{\omega-k_z v_m}-\frac{v_f}{\omega-k_z v_f}\bigg)e^{-i(\omega-k_z v_m)T/2}\bigg]~{\bf e_z}.~~
\label{2stepsj}
\end{eqnarray}

If the intermediate time interval $T$ is extremely short, Eq.~(\ref{2stepsj}) returns to Eq.~(\ref{stepj1}),

\begin{eqnarray}
\lim_{T\rightarrow 0}{\bf j}(\omega,{\bf k})=iQ\bigg(\frac{v_i}{\omega-k_z v_i} - \frac{v_f}{\omega-k_z v_f}\bigg)~{\bf e_z}.
\end{eqnarray}

We can interpret Eq.~(\ref{2stepsj}) as following:
The first and second terms represent the first and second photon emissions at $t=-T/2$ and $t=T/2$, respectively, and the two exponential functions show their phases.
Therefore, Eq.~(\ref{2stepsj}) can be generalized to N-photon emissions~\cite{Cleymans:1992kb},

\begin{eqnarray}
{\bf j}(w,{\bf k})=iQ\sum_{i=1}^N \bigg(\frac{v_{i-}}{w-k_z v_{i-}} - \frac{v_{i+}}{w-k_z v_{i+}}\bigg)e^{-i\phi_{i1}}~{\bf e_z},~~
\label{nstepsj}
\end{eqnarray}
where $v_{i-}$ and $v_{i+}$ are respectively velocities of charged particle before and after the $i-$th photon emission, and $\phi_{i1}$ phase difference between the first and the $i-$th photons, which is given by
\begin{eqnarray}
\phi_{11}&=&0,\\
\phi_{i1}&=&\sum_{j=1}^{i-1}(\omega-k_z v_{j+})\Delta t_j^{j+1}\nonumber\\
&=&\sum_{j=1}^{i-1} \omega(1- v_{j+}\cos\theta)\Delta t_j^{j+1}
\end{eqnarray}
with $\Delta t_j^{j+1}$ being the time interval between the $j-$th and the $j+1-$th photon emissions.
If the photon energy $\omega$ is small enough, we may ignore the phase differences in Eq.~(\ref{nstepsj}), and the electromagnetic current turns out to be

\begin{eqnarray}
\lim_{\omega\rightarrow 0}{\bf j}(\omega,{\bf k})=iQ\sum_{i=1}^N \bigg(\frac{v_{i-}}{\omega-k_z v_{i-}} - \frac{v_{i+}}{\omega-k_z v_{i+}}\bigg)~{\bf e_z}\nonumber\\
=iQ \bigg(\frac{v_i}{\omega-k_z v_i} - \frac{v_f}{\omega-k_z v_f}\bigg)~{\bf e_z},~~~
\end{eqnarray}
which is equivalent to Eq.~(\ref{stepj1}).
It explains why the photon spectrum near $\omega=0$ does not change for various stopping times in figure~\ref{smooth1} (b).
Low-energy photon can not provide the information of short time scale.

Now we apply the stepwise method to the previous subsection where a charged particle moves with a constant velocity but electric charge evaporates with time.
Supposing electric charge changes from $Q_i$ to $Q_f$ at $t=0$,
electromagnetic current is given by

\begin{eqnarray}
{\bf j}(t,{\bf r})=Q(t)v\delta(z-vt)\delta(x)\delta(y)~{\bf e_z},
\end{eqnarray}
where
\begin{eqnarray}
Q(t)=Q_i\theta(-t) +Q_f\theta(t),
\end{eqnarray}

and after Fourier transformations it turns to
\begin{eqnarray}
{\bf j}(\omega,{\bf k})=\int dt d^3{\bf r} {\bf j}(t,{\bf r})e^{i({\bf k}\cdot {\bf r}-\omega t)}\nonumber\\
=i\frac{(Q_i-Q_f)v}{\omega-k_zv}~{\bf e_z}.
\label{stepj2}
\end{eqnarray}

Next suppose particle changes its electric charge from $Q_i$ to $Q_m$ at $t=-T/2$, and then from $Q_m$ to $Q_f$ at $t=T/2$:
\begin{eqnarray}
Q(t)=Q_i\theta(-t-T/2)+Q_m{\rm rect}_T(t) +Q_f\theta(t-T/2).\nonumber\\
\end{eqnarray}

Then Fourier-transformed current is given by
\begin{eqnarray}
{\bf j}(\omega,{\bf k})=\int dt d^3{\bf r} {\bf j}(t,{\bf r})e^{i({\bf k}\cdot {\bf r}-\omega t)}\nonumber\\
=iv\bigg\{ \frac{Q_i-Q_m}{\omega-k_z v}e^{-i(k_z v-\omega)T/2}~~~~~\nonumber\\
+\frac{Q_m-Q_f}{\omega-k_z v}e^{i(k_z v-\omega)T/2} \bigg\}~{\bf e_z}.
\end{eqnarray}
Compared to Eq.~(\ref{stepj2}), it is nothing but the summation of two currents with phase terms.
Therefore we can generalize it to N-photon emissions as before:

\begin{eqnarray}
{\bf j}(\omega,{\bf k})=iv\sum_{i=1}^N \frac{Q_{i-}-Q_{i+}}{\omega-k_z v} e^{-i\phi_{i1}}~{\bf e_z},
\label{nstepj2}
\end{eqnarray}
where $Q_{i-}$ and $Q_{i+}$ are respectively particle charges before and after the $i-$th photon emission and $\phi_{i1}$ phase difference between the first and the $i-$th photons, which is given by
\begin{eqnarray}
\phi_{11}=0,\\
\phi_{i1}=\sum_{j=1}^{i-1}(\omega-k_z v)\Delta t_j^{j+1}\nonumber\\
=\sum_{j=1}^{i-1} \omega(1- v\cos\theta)\Delta t_j^{j+1}
\end{eqnarray}
with $\Delta t_j^{j+1}$ being the time interval between the $j-$th photon emission and the $j+1-$th photon emission.
We point out that at mid-rapidity ($\cos\theta=0$) Eqs.~~(\ref{nstepsj}) and (\ref{nstepj2}) are expressed in unified form,

\begin{eqnarray}
{\bf j}(\omega,{\bf k})=i\sum_{i=1}^N \frac{j_{i-}-j_{i+}}{\omega} e^{-i\phi_{i1}}~{\bf e_z},
\label{nsteps3}
\end{eqnarray}
where $j_{i-}\equiv Q_{i-}v_{i-}$ and $j_{i+}\equiv Q_{i+}v_{i+}$ are respectively electromagnetic currents before and after the $i-$th photon emission with the phase difference
\begin{eqnarray}
\phi_{11}=0,~~~~~\phi_{i1}=\sum_{j=1}^{i-1}\omega\Delta t_j^{j+1}.
\end{eqnarray}

In other words, Bremsstrahlung photon from decelerated charged particle and that from the particle whose electric charge evaporates are indistinguishable at mid-rapidity, if the particle velocity and the evaporation speed are same.

\begin{figure} [h]
\centerline{
\includegraphics[width=8.6 cm]{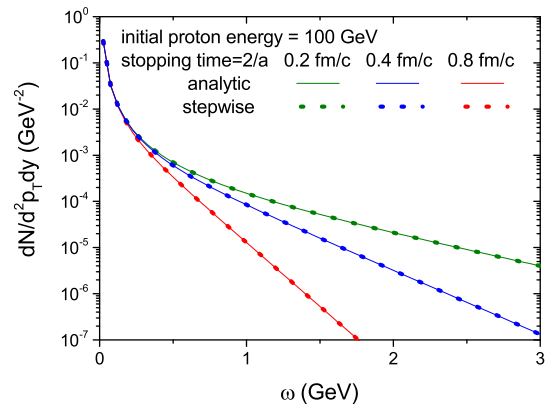}}
\caption{(Color online) photon spectra at mid-rapidity from proton stopping with initial energy of 100 GeV between analytic solution in Eq.~(\ref{smoothg}) and the results from the stepwise method} \label{resolution}
\end{figure}

Figure~\ref{resolution} compares photon spectra at mid-rapidity from proton stopping with initial energy of 100 GeV between analytic solution in Eq.~(\ref{smoothg}) and the results from the stepwise method.
We can see that the stepwise method reproduces the analytic solution.

\section{consideration of nucleon structure}
\label{structure}

Proton is not a point-like particle but a composite particle made up of at least three valence quarks.
In this section we substitute proton with three valence quarks which are randomly distributed in a sphere with radius $R$.
Since the photon spectra from charge evaporation and from deceleration are same at mid-rapidity, we take the former case for simplicity.
Electromagnetic current from three comoving valence quarks is given by

\begin{eqnarray}
{\bf j}(t,{\bf r})=\sum_{i=1\sim 3} Q_iv\delta(z-vt)\delta(x-x_i)\delta(y-y_i)\nonumber\\
\times \frac{1-\tanh(at)}{2}~{\bf e_z},
\end{eqnarray}
where $Q_i$ is the electric charge of valence quark $i$ and $z_i=0$ at $t=0$ is assumed from the Lorentz contraction in ultrarelativistic collisions.
We also assume that $x_i$ and $y_i$ do not change with time, even after collision.
The current is transformed in momentum space as followings:

\begin{eqnarray}
{\bf j}(t,{\bf k})=\int d^3{\bf r} {\bf j}(t,{\bf r})e^{i{\bf r\cdot k}}\nonumber\\
=\sum_{i=1\sim 3} Q_i v \frac{1-\tanh(at)}{2}e^{i(k_xx_i+k_yy_i+k_z vt)}~{\bf e_z},
\end{eqnarray}
and

\begin{eqnarray}
{\bf j}(\omega,{\bf k})=\int dt {\bf j}({\bf k},t)e^{-i\omega t}\nonumber\\
=i\frac{\pi v}{2a}{\rm csch}\bigg(\frac{\pi (\omega-k_zv)}{2a}\bigg)\sum_{i=1\sim 3} Q_ie^{i(k_xx_i+k_yy_i)}~{\bf e_z}.
\label{jquarks}
\end{eqnarray}

Therefore we can find the relation,
\begin{eqnarray}
\omega\frac{dN}{d^3k}\bigg|_{\rm~ with~structure}=\omega\frac{dN}{d^3k}\bigg|_{\rm~ point-like}\nonumber\\
\times\bigg|\sum_{i=1\sim 3} q_ie^{i(k_xx_i+k_yy_i)}\bigg|^2,
\end{eqnarray}
where $q_i=Q_i/(\sum_{j=1\sim 3} Q_j)$.
Rotating the coordinate system such that $(k_x,~k_y)\rightarrow (k_T,~0)$, the correction factor is simplified into
\begin{eqnarray}
\bigg|\sum_{i=1\sim 3} q_ie^{i(k_xx_i+k_yy_i)}\bigg|^2
=\sum_{i,j=1\sim 3}q_iq_je^{ik_T( x_i-x_j)}\nonumber\\
=\sum_{i,j=1\sim 3}q_iq_j\cos \{k_T( x_i-x_j)\},
\end{eqnarray}
where sine term in the last equation vanishes.
In the limit $k_T\rightarrow 0$, the correction factor turns to unity:
\begin{eqnarray}
\lim_{k_T\rightarrow 0} \bigg|\sum_{i=1\sim 3} q_ie^{i(k_xx_i+k_yy_i)}\bigg|^2=\sum_{i,j=1\sim 3}q_iq_j=1.
\end{eqnarray}
It means again that photon with a very small frequency does not provide the information of detailed structure of charged particle.
We can calculate the expectation value of cosine function in the case $x_i\neq x_j$ as following:

\begin{eqnarray}
&&\langle \cos \{k_T( x_i-x_j)\} \rangle\bigg|_{x_i\neq x_j} =\frac{1}{(4/3\pi R^3)^2}\int dV_i dV_j\nonumber\\
&&\times\{\cos (k_T x_i)\cos(k_T x_j)+\sin(k_T x_i)\sin(k_T x_j)\}\nonumber\\
&&=\frac{1}{(2/3 R^3)^2}\bigg\{\int_0^R dr_i r_i^2 \int_{-1}^1 d\cos\theta_i \cos (k_T r_i\cos\theta_i)\bigg\}^2\nonumber\\
&&=\frac{9}{(k_T R)^6}\{-k_TR\cos(k_TR)+\sin(k_TR)\}^2,
\label{interference}
\end{eqnarray}
where $\theta_i$ is the angle between ${\bf k_T}$ and ${\bf r_i}$, and finally
\begin{eqnarray}
\bigg|\sum_{i=1\sim 3} q_ie^{i(k_xx_i+k_yy_i)}\bigg|^2=\sum_{i=1\sim 3}q_i^2\nonumber\\
+\frac{9}{(k_T R)^6}\{-k_TR\cos(k_TR)+\sin(k_TR)\}^2\sum_{i\neq j}q_iq_j.
\end{eqnarray}

\begin{figure}[h]
\centerline{
\includegraphics[width=8.6 cm]{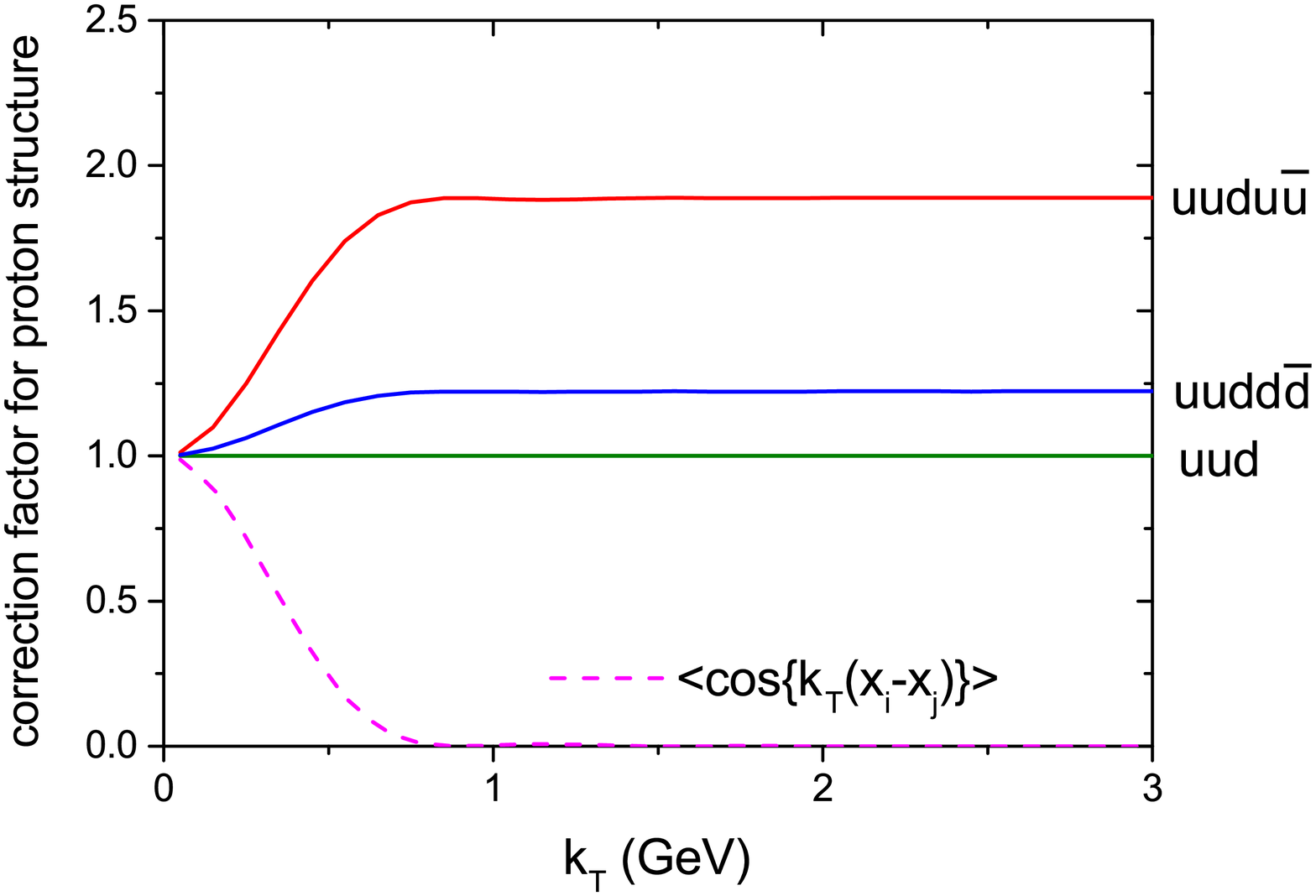}}
\caption{(Color online) correction factors for proton structure as a function of transverse momentum of photon for different combinations of quarks ($uud,~uudu\bar{u},~uudd\bar{d}$). Proton radius $R$ is taken to be 1 fm, and the magenta line shows Eq.~(\ref{interference}).}
\label{structure2}
\end{figure}

Figure~\ref{structure2} shows the correction factors for proton structure as a function of transverse momentum of photon with proton radius $R$ being taken to be 1 fm.
In order to get an insight we use three different combinations of quarks ($uud,~uudu\bar{u},~uudd\bar{d}$) for proton.
The magenta line shows Eq.~(\ref{interference}), which starts with 1.0 at $k_T=$ 0 GeV and then almost vanishes before $k_T=$ 1 GeV.
It explains the behavior of correction factors in figure~\ref{structure2}.
For example, supposing that proton is composed of $uudu\bar{u}$, the correction factor at small $k_T$

\begin{eqnarray}
\bigg(\frac{2}{3}+\frac{2}{3}-\frac{1}{3}+\frac{2}{3}-\frac{2}{3}\bigg)^2=1,
\end{eqnarray}
and at large $k_T$
\begin{eqnarray}
\bigg(\frac{2}{3}\bigg)^2+\bigg(\frac{2}{3}\bigg)^2+\bigg(-\frac{1}{3}\bigg)^2+\bigg(\frac{2}{3}\bigg)^2+\bigg(-\frac{2}{3}\bigg)^2=\frac{17}{9},\nonumber\\
\end{eqnarray}

which means that individual quark is not seen at small $k_T$ but seen at large $k_T$.
In other words, coherent photon smoothly changes into incoherent photon, as photon energy increases.
However, if only valence quarks are considered ($uud$), there is no difference between coherent and incoherent photons, because the correction factor

\begin{eqnarray}
\bigg(\frac{2}{3}+\frac{2}{3}-\frac{1}{3}\bigg)^2=1
\end{eqnarray}
at small $k_T$ is equivalent to
\begin{eqnarray}
\bigg(\frac{2}{3}\bigg)^2+\bigg(\frac{2}{3}\bigg)^2+\bigg(-\frac{1}{3}\bigg)^2=1
\end{eqnarray}
at large $k_T$.
We may consider proton as the combination of $uudu\bar{u}$ or $uudd\bar{d}$, taking $u\bar{u}$ or $d\bar{d}$ for sea quark pair.
In this case, however, sea quark and sea antiquark are not located independently in nucleon but strongly correlated in space so that their effect will appear at much larger photon energy.

\section{stopping in collision}
\label{collision}

In this section we describe the stopping of two particles in collision.
For simplicity two particles move in opposite directions with the same velocity and electric charges, $Q_1$ and $Q_2$, evaporate with time:
\begin{eqnarray}
{\bf j}(t,{\bf r})=v\delta(y)\frac{1-\tanh(at)}{2}\{Q_1\delta(x-b/2)\delta(z-vt)\nonumber\\
-Q_2\delta(x+b/2)\delta(z+vt)\}~{\bf e_z},
\end{eqnarray}
where $b$ is the impact parameter~\cite{Koide:2016kpe}.
Taking Fourier transformations,

\begin{eqnarray}
{\bf j}(t,{\bf k})=\int d^3{\bf r} {\bf j}(t,{\bf r})e^{i{\bf r\cdot k}}=v \frac{1-\tanh(at)}{2}\nonumber\\
\times\{Q_1e^{i(k_z vt+k_x b/2)}-Q_2e^{-i(k_z vt+k_x b/2)}\}~{\bf e_z},
\end{eqnarray}
and

\begin{eqnarray}
{\bf j}(\omega,{\bf k})&=&\int dt {\bf j}(t,{\bf k})e^{-i\omega t}\nonumber\\
&=&i\frac{\pi v}{2a}\bigg\{ Q_1{\rm csch}\bigg(\frac{\pi (\omega-k_zv)}{2a}\bigg)e^{ik_x b/2} \nonumber\\
&&-Q_2{\rm csch}\bigg(\frac{\pi (\omega+k_zv)}{2a}\bigg)e^{-ik_x b/2}\bigg\}~{\bf e_z}.
\label{impactj}
\end{eqnarray}
The only difference of Eq.~(\ref{impactj}) from Eq.~(\ref{smoothj}) is two phase terms which is ascribed to the separation of two currents in x-direction.
It is straightforward to prove that the photon spectrum from the evaporation of two electric charges and that from the deceleration of two charged particles are same at mid-rapidity as shown in the previous section.
By using Eq.~(\ref{impactj}) photon spectrum at mid-rapidity ($k_z=0$) from the collision turns out

\begin{eqnarray}
\frac{dN}{d^2k_T dy}\bigg|_{y=0}
=\frac{1}{2(2\pi)^3}\sum_\lambda| {\bf j}(\omega,{\bf k})\cdot {\bf \epsilon}^\lambda(\omega,{\bf k})|^2\nonumber\\
=\frac{1}{16\pi}\bigg(\frac{v}{2a}\bigg)^2 {\rm csch}^2\bigg(\frac{\pi \omega}{2a}\bigg)\bigg\{(Q_1-Q_2)^2\nonumber\\
+4Q_1Q_2\sin^2\bigg(\frac{k_x b}{2}\bigg)\bigg\}.
\label{impacts}
\end{eqnarray}

Eq.~(\ref{impacts}) applied to p+n collisions ($Q_1=e,~Q_2=0$),

\begin{eqnarray}
\frac{dN}{d^2k_T dy}\bigg|_{y=0}=\frac{1}{16\pi}\bigg(\frac{j}{2a}\bigg)^2{\rm csch}^2\bigg(\frac{\pi \omega}{2a}\bigg),
\label{pn-mid}
\end{eqnarray}
and to p+p collisions ($Q_1=Q_2=e$),
\begin{eqnarray}
\frac{dN}{d^2k_T dy}\bigg|_{y=0}=\frac{1}{4\pi}\bigg(\frac{j}{2a}\bigg)^2{\rm csch}^2\bigg(\frac{\pi \omega}{2a}\bigg)\sin^2\bigg(\frac{k_x b}{2}\bigg),
\label{pp-mid}
\end{eqnarray}
where $j=Q_iv_i$.
We note that if the impact parameter $b=0$ in p+p collision, photon spectrum vanishes at mid-rapidity, though it hardly happens in reality.
If electric charges do not completely stop, which is usual in high-energy collisions, the electromagnetic currents $j$ in Eqs.~(\ref{pn-mid}) and (\ref{pp-mid}) are substituted by $\Delta j=Q_iv_i-Q_fv_f$, the change of electromagnetic current.

Eq.~(\ref{impactj}) can be expressed in terms of valence quark, substituting $Q_1$ and $Q_2$ by $\sum_j Q_je^{i{\rm k}\cdot {\rm r}_j}$ as in Eq.~(\ref{jquarks}):

\begin{eqnarray}
{\bf j}(\omega,{\bf k}_T,k_z=0)=i\frac{\pi v}{2a}{\rm csch}\bigg(\frac{\pi \omega}{2a}\bigg)~~~~~~~~~~~~~~~\nonumber\\
\times\sum_i \bigg(Q_{1i}e^{ik_T\cdot ({\bf r}_{1i}+{\bf b}/2)}-Q_{2i}e^{ik_T\cdot ({\bf r}_{2i}-{\bf b}/2)}\bigg)~{\bf e_z},\nonumber\\
\end{eqnarray}
where $Q_{1i}(Q_{2i})$ and ${\bf r}_{1i}({\bf r}_{2i})$ are, respectively, the electric charge and the transverse position from the center of nucleon 1(2) of quark $i$, which composes nucleon 1(2).
Following the previous section it is straightforward to calculate photon spectrum:

\begin{eqnarray}
\frac{dN}{d^2k_T dy}\bigg|_{y=0}=\frac{1}{16\pi}\bigg(\frac{v}{2a}\bigg)^2 {\rm csch}^2\bigg(\frac{\pi \omega}{2a}\bigg)\nonumber\\
\times \bigg[\sum_{i,j}Q_{1i}Q_{1j}\cos \{k_T( x_{1i}-x_{1j})\}\nonumber\\
+\sum_{i,j}Q_{2i}Q_{2j}\cos \{k_T( x_{2i}-x_{2j})\}\nonumber\\
-2\sum_{i,j}Q_{1i}Q_{2j}\cos \{k_T( x_{1i}-x_{2j}+b\cos\phi)\}\bigg],
\end{eqnarray}
where coordinate system is rotated such that $k_T$ is parallel to ${\bf e_x}$ and $\phi$ is the angle between $b$ and $k_T$.
Assuming quarks are randomly distributed in nucleons whose radius is $R$,

\begin{eqnarray}
\frac{dN}{d^2k_T dy}\bigg|_{y=0}=\frac{1}{16\pi}\bigg(\frac{v}{2a}\bigg)^2 {\rm csch}^2\bigg(\frac{\pi \omega}{2a}\bigg)~~~~~\nonumber\\
\times \bigg[\sum_{i}(Q_{1i}^2 +Q_{2i}^2)~~~~~~~~~~~~~~~~~~~~~~~~~~~~~~\nonumber\\
+\frac{9}{(k_T R)^6}\{-k_TR\cos(k_TR)+\sin(k_TR)\}^2\nonumber\\
\times \bigg\{\sum_{i\neq j}(Q_{1i}Q_{1j}+Q_{2i}Q_{2j})~~~~~~~~\nonumber\\
-2\cos(k_Tb\cos\phi)\sum_{i j}Q_{1i}Q_{2j}\bigg\}\bigg],
\label{sp-quarks}
\end{eqnarray}
which converges to Eq.~(\ref{impacts}) in the limit $R\rightarrow 0$.

In the picture of pQCD, only one parton in nucleon interacts with one parton from the other nucleon.
However, if a valence quark gets out of nucleon by the scattering, the remaining two valence quarks cannot proceed without interaction, because they are not color-singlet any more.
They should somehow be involved in the scattering.
Though the stopping or deceleration of three valence quarks might be different from each other, we can simply take their average.

We note that Eq.~(\ref{sp-quarks}) can be applied to nucleus-nucleus collision with $Q_{1i}(Q_{2i})$ and $R$ being respectively electric charge of nucleon $i$ in nucleus 1(2) and nucleus radius, if all nucleons are participants.
In the case of incomplete stopping, $vQ_{1i}(vQ_{2i})$ is replaced by $\Delta j_{1i}(\Delta j_{2i})$.

\section{relativistic N+N collisions}
\label{experiment}

If collision energy is low, nucleon+nucleon scattering would be elastic or excitation such as $N+N\rightarrow N+\Delta$.
However, when the collision energy is extremely large, two colliding nucleons pass through each other and only part of energy and electromagnetic current are released, which produces both charged and neutral particles.
Since wounded nucleons go straight even after collision and each nucleon has a finite size, we can approximate it to one-dimensional stopping or deceleration of electromagnetic currents as in the previous section.

For two reasons it is hard to know how much fraction of electromagnetic current is stopped in high-energy nucleon-nucleon collisions.
Firstly, it is experimentally challenging to measure particles in very large rapidity regions.
Secondly, the charge stopping is not well defined in the collision of equal-charged particles such as p+p collisions, since total electromagnetic current of the system is zero.

\begin{figure} [h!]
\centerline{
\includegraphics[width=8.6 cm]{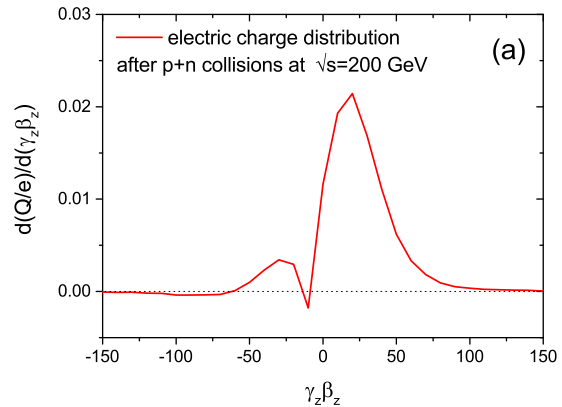}}
\centerline{
\includegraphics[width=8.6 cm]{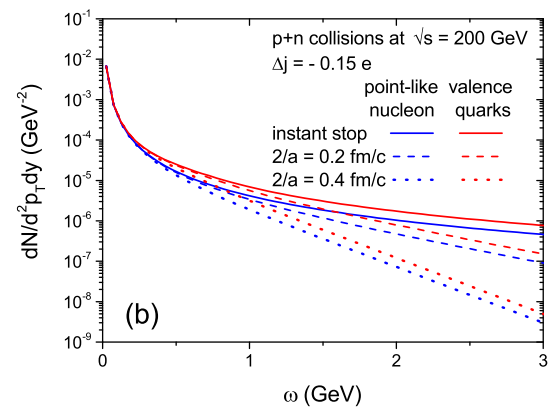}}
\centerline{
\includegraphics[width=8.6 cm]{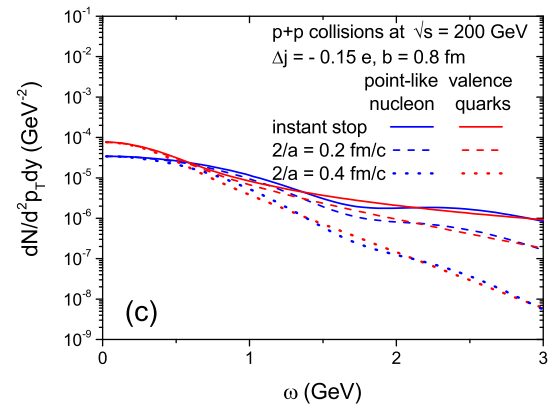}}
\caption{(Color online) (a) electric charge distribution as a function of $\gamma_z\beta_z=\beta_z/\sqrt{1-\beta_z^2}$ in p+n collisions at $\sqrt{s}=$ 200 GeV from the PYTHIA event generator, and photon spectra for different stopping times in (b) p+n and (c) p+p collisions at the same collision energy, assuming a point-like particle or a group of valence quarks for nucleon.} \label{nn}
\end{figure}

We use the PYTHIA event generator to solve the first problem~\cite{Sjostrand:2006za}.
The upper panel of figure~\ref{nn} shows the electric charge distribution as a function of $\gamma_z\beta_z=\beta_z/\sqrt{1-\beta_z^2}$ in p+n collisions at $\sqrt{s}=$ 200 GeV from the PYTHIA event generator, where proton moves in +z direction and neutron in the opposite.
Electromagnetic current after the collision is 0.85 $e$ on average and we thus take $\Delta j=$ -0.15 $e$.
The middle panel shows Bremsstrahlung photon spectra for different stopping times in p+n collisions by using Eqs.~(\ref{pn-mid}) and (\ref{sp-quarks}), assuming a point-like particle or a group of valence quarks for nucleon.
We note that the impact parameter in the two equations does not affect the spectrum in p+n collisions, since $Q_2=$0.
In both cases photon spectrum becomes soft with increasing stopping or deceleration time, while the spectrum at low $\omega$ does not change, which is consistent with figure~\ref{resolution}.
We can see that photon spectrum is larger at high $\omega$ in the case of three valence quarks compared to that for point-like nucleon.
If neutron is a point-like particle, it cannot emit photon, while three valence quarks of it ($udd$) can emit photons respectively in spite of interferences, which are completely destructive at low $\omega$ but become incoherent at high $\omega$, as shown in figure~\ref{structure2}. That is the reason for the enhancement of photon spectra at high $\omega$ in the case of three valence quarks.

In p+p collisions, total electromagnetic current vanishes.
Furthermore, it is not clear whether a charged particle produced in the collision is originated from the target proton or from the projectile proton.
We assume that the same amount of electromagnetic current is stopped in p+n and p+p collisions.
The lower panel of figure~\ref{nn} shows the photon spectra at mid-rapidity in p+p collisions at $\sqrt{s}=$ 200 GeV, assuming point-like nucleons or two groups of three valence quarks.
Considering the inelastic scattering cross section of 42 mb, impact parameter for minimum-bias events is about 0.8 fm on average.
As in p+n collisions, the spectrum of Bremsstrahlung becomes soft with increasing stopping or deceleration time, while the spectrum at low $\omega$ does not change.
Since proton has the same contribution whether it is a point-like particle or a group of three valence quarks, as shown in figure~\ref{structure2}, two different pictures bring about similar photon spectra except the fluctuations which are ascribed to the interference of photons from target and projectile protons. The fluctuations are more prominent in the case of point-like protons.

Here we point out that the spectrum of Bremsstrahlung photon at low energy in p+p collisions is considerably different from that in p+n collisions.
Therefore, if photon energy is low, it is not right to scale photon spectrum only in p+p collisions to extract nuclear matter effect in heavy-ion collisions.
The number of binary collisions should be separated into the number of p+p collisions, that of p+n collisions, and that of n+n collisions.

\begin{figure} [h]
\centerline{
\includegraphics[width=8.6 cm]{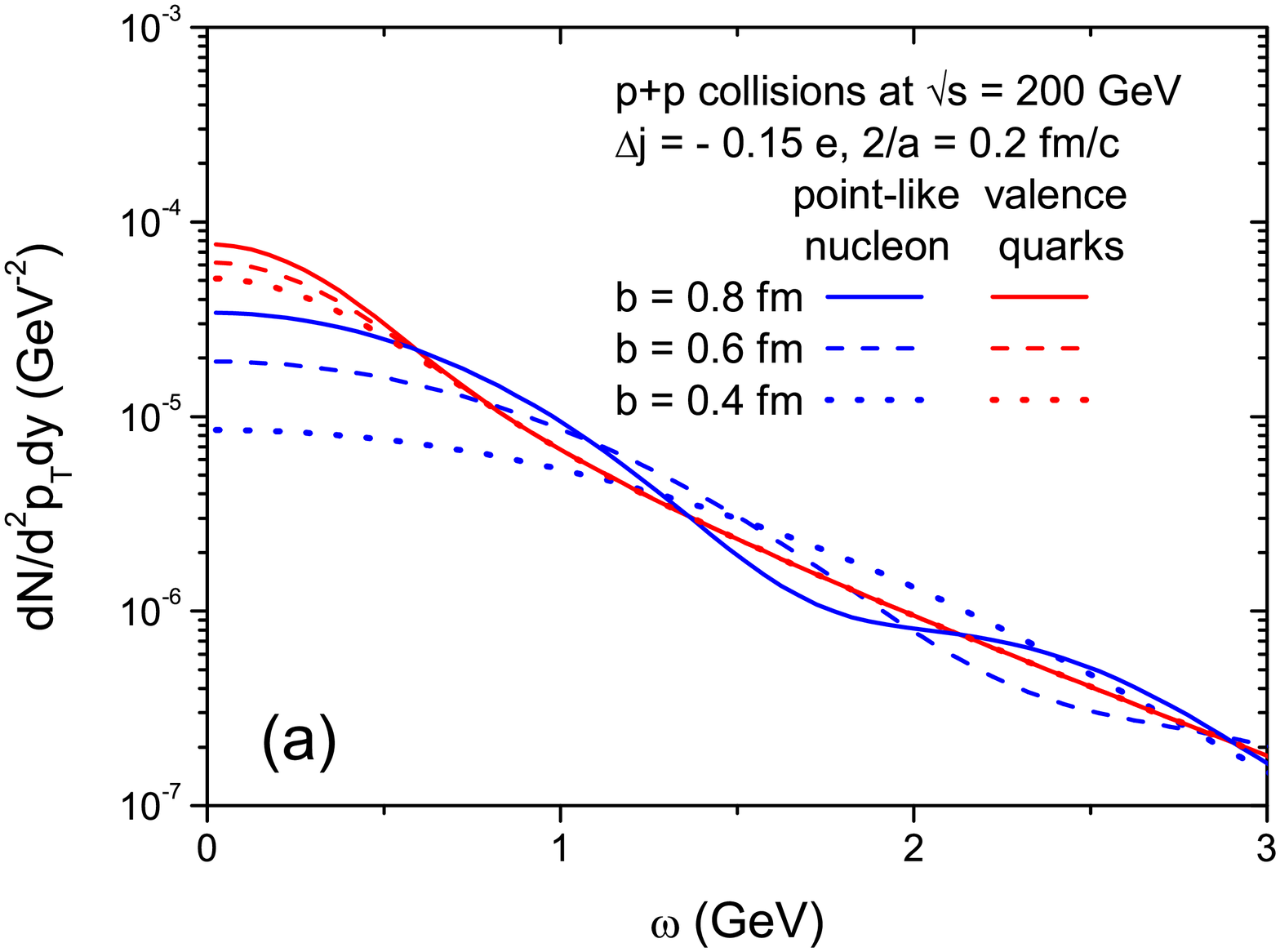}}
\centerline{
\includegraphics[width=8.6 cm]{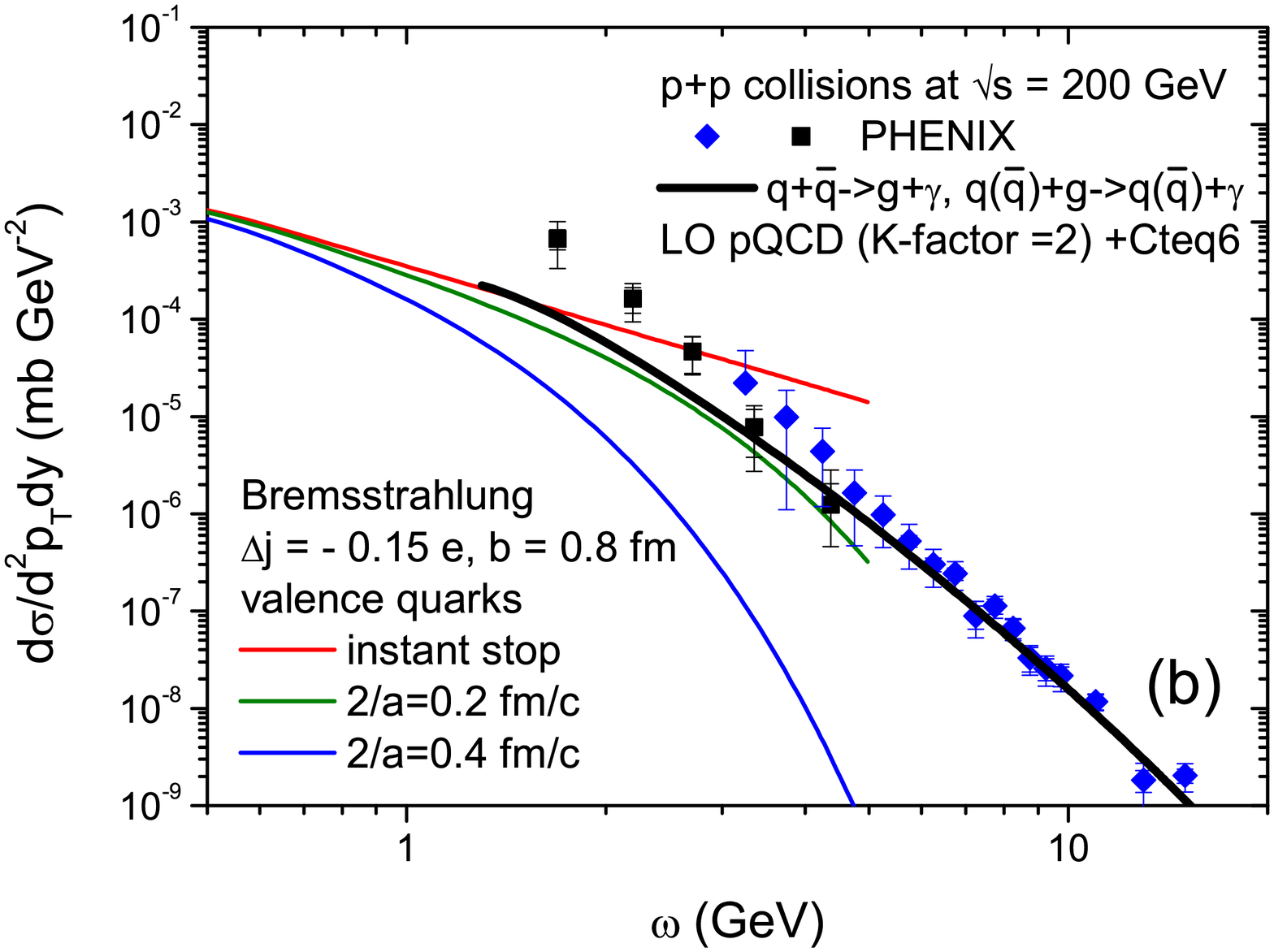}}
\caption{(Color online) (a) spectra of Bremsstrahlung photon at mid-rapidity in p+p collisions at $\sqrt{s}=$ 200 GeV for the stopping time of 0.2 fm/c and a couple of separation distances and (b) the spectra for $\rm b=$ 0.8 fm and a couple of stopping times in comparison with the experimental data on direct photon from the PHENIX Collaboration~\cite{Adler:2006yt,Adare:2008ab}, along with pQCD calculations.} \label{pp-fig}
\end{figure}

The impact parameter in Eqs.~(\ref{impacts}) and (\ref{sp-quarks}) does not necessarily mean geometric impact parameter but an effective distance between stopped charges of two colliding protons.
Since it will be smaller than the geometric impact parameter, we try not only with 0.8 fm but also with smaller ones in the upper panel of figure~\ref{pp-fig}.
It shows that photon spectrum is not sensitive to the separation distance especially in the case of three valence quarks where little differences are seen only below $\omega=$ 0.5 GeV.

The lower panel shows the experimental data on direct photon in p+p collisions at $\sqrt{s}=$ 200 GeV from the PHENIX Collaboration~\cite{Adler:2006yt,Adare:2008ab}, which are compared with pQCD calculations to the leading-order for $q+\bar{q}\rightarrow g+\gamma$ and $q(\bar{q})+g\rightarrow q(\bar{q})+\gamma$~\cite{Wong:1998pq} with the CTEQ parton distribution function~\cite{Pumplin:2002vw}.
The scale of parton distribution function is set at photon energy and the K-factor, which takes into account higher-order corrections, is taken to be 2.
We can see that the pQCD calculations reproduce the experimental data down to $\omega=3-4$ GeV and then deviate from them at lower photon energy.
Since the calculations are not reliable at low photon energy, they are shown only down to $\omega=$ 1.3 GeV in the figure.

We also show Bremsstrahlung photon for $\rm b=$ 0.8 fm and 0, 0.2, and 0.4 fm/c of the stopping time in Eq.~(\ref{sp-quarks}).
Though the comparison with the experimental data cannot say something definite,
it seems that the deceleration of electromagnetic current in p+p collision does not take place instantly but takes a time longer than 0.2 fm/c.

Naively thinking, the deceleration time would be the duration when two nucleons pass through each other.
Considering Lorentz contraction, it is only a couple of hundredth fm/c.
However, rich interactions are still left after that, for example, string fragmentation, particle production, and so on.
Though the deceleration of electromagnetic currents in p+p collisions may not follow the pattern of figure~\ref{smooth1}, it would be possible to estimate the deceleration time from Bremsstrahlung photon spectrum.

\section{summary}
\label{summary}

Relativistic heavy-ion collisions produce hot dense nuclear matter.
Photon is a clear probe for the properties of the nuclear matter, because it hardly interacts after production.
Since photon is produced from the initial stage to the final one in heavy-ion collisions, it needs to be classified according to when it is produced.

At first nucleons composing nuclei lose considerable energy and momentum through their primary collisions, which bring about the radiation of photon.
In microscopic picture photon is produced through the scattering of partons in nucleons.
The collision of heavy nuclei produces extremely hot nuclear matter, and the matter cools down by emitting thermal photons.
Finally the matter freeze out as noninteracting hadrons, which produce photons through electromagnetic decay.
The photon excluding the last case is called direct photon.

Nucleon+nucleon collision is a reference experiment to extract nuclear matter effect from heavy-ion collisions, because it hardly produces a sizeable matter.
Since nucleon+nucleon collisions do not produce thermal photons, all photons excluding those from electromagnetic decay are direct photon, which is produced mostly in the initial stage of collisions.
The production of direct photon can be described by parton interactions in pQCD combined with parton distribution function, if the photon energy is large enough.
However, pQCD does not work for low-energy photon.

In this paper we have studied the production of Bremsstrahlung photon in relativistic nucleon+nucleon collisions, which is not restricted to high-energy photon but applicable also to low-energy one.
Since nucleon is not an elementary particle but a composite particle with structure, the stopping of electromagnetic current is modeled by hyperbolic tangent function with a parameter for stopping time.
We also approximate the collisions to one-dimensional stopping of electromagnetic currents, because two nucleons pass through each other in high-energy collisions.
In general, it is hard to measure the amount of stopped electromagnetic current in collisions, because all range of rapidity should be covered by detectors.
Therefore we use the PYTHIA event generator and have found that about 15 \% of initial electromagnetic current stops in p+n collisions at $\sqrt{s}=$ 200 GeV.
The same amount of stopping is assumed in p+p collisions.

We have found that Bremsstrahlung photon spectrum at low energy does not depend on stopping or deceleration time but only on the amount of stopped electromagnetic current, because low energy photon cannot provide the information of short time scale.
Beyond this energy range, however, the spectrum becomes soft with increasing stopping time.

We have also studied the effect of nucleon structure by substituting comoving three valence quarks for point-like nucleon.
In p+n collisions, the substitution enhances photon spectrum at large energy, because neutron cannot emit photon while three valence quarks composing the neutron can emit photons respectively, and
each photon becomes incoherent as photon energy increases.
In p+p collisions Bremsstrahlung photon can be emitted from both projectile and target protons and interference could be destructive or constructive.
For example, if the impact parameter in p+p collision vanishes, the interference is completely destructive and no Bremsstrahlung photon is produced.
In the picture of comoving valence quarks, however, the fluctuations of Bremsstrahlung photon spectrum caused by the interference reduce, and the photon does not vanishes even at $b=$0.

The photon spectrum in p+n collisions and that in p+p collisions are significantly different from each other at low energy, and we suggest that the photon spectrum in p+p collisions should not be rescaled simply by the number of binary collisions in heavy-ion collisions, when nuclear matter effect is studied, but the number of binary p+p collisions, that of p+n collisions, and that of n+n collisions should be separately counted.

Comparing the Bremsstrahlung photon spectrum with the experimental data on direct photon from the PHENIX Collaboration, it seems that
the stopping of electromagnetic current in relativistic p+p collisions does not take place instantly but takes a time, which is longer than the overlapping time of two nucleons, because there are additional processes such as particle production.

\section*{Acknowledgements}
The authors acknowledge inspiring discussions with W. Cassing, and E.
Bratkovskaya.
This work was supported by the LOEWE center "HIC for FAIR", the HGS-HIRe for FAIR and
the COST Action THOR, CA15213.
Furthermore, PM and EB acknowledge support by DFG through the grant CRC-TR 211 'Strong-interaction matter under extreme conditions'.
The computational resources have been provided by the LOEWE-CSC.


\hfil\break
\appendix
\bigskip

\section{}
In this appendix we show the Fourier transformation of basic functions.

\subsection{signum function}
The signum function is defined as
\begin{eqnarray}
{\rm sgn}(t)=1 ~(t \geq 0),\nonumber\\
=-1 ~(t < 0).
\end{eqnarray}
It can be expressed as
\begin{eqnarray}
{\rm sgn}(t)=\lim_{a\rightarrow 0^+}\{e^{-at}\theta(t)-e^{at}\theta(-t)\},
\end{eqnarray}
where $\theta(t)$ is the step function and the Fourier transformation of the signum function is derived as below:
\begin{eqnarray}
\textit{F}_w[{\rm sgn}(t)]=\lim_{a\rightarrow 0^+}\bigg(\int_0^\infty e^{-at}e^{-iwt}dt-\int_{-\infty}^0 e^{at}e^{-iwt}dt\bigg)\nonumber\\
=\lim_{a\rightarrow 0^+}\bigg(\frac{1}{a+iw}-\frac{1}{a-iw}\bigg)=\frac{2}{iw}.~~~~~
\label{signum}
\end{eqnarray}

\subsection{step function}
The step function is expressed by using the signum function
\begin{eqnarray}
\theta(t)=\frac{1}{2}\{1+{\rm sgn}(t)\},
\end{eqnarray}
and after Fourier transformation
\begin{eqnarray}
\textit{F}_w[\theta(t)]=\frac{1}{2}\{\textit{F}_w[1]+\textit{F}_w[{\rm sgn}(t)]\}=\pi\delta(w)+\frac{1}{iw}.
\label{stepf}
\end{eqnarray}

\subsection{box function}
Supposing the box function is centered at $t=0$ with the width $T$
\begin{eqnarray}
{\rm rect}_T(t)&=&1 ,~{\rm if}~ |t|\leq T/2,\nonumber\\
{\rm rect}_T(t)&=&0 ,~{\rm if}~ |t|> 0,
\label{boxf}
\end{eqnarray}

then the Fourier transformation is carried out as following:
\begin{eqnarray}
\int {\rm rect}_T(t)e^{-iwt}dt= \int_{-T/2}^{T/2} e^{-iwt}dt\nonumber\\
=\frac{i}{w}(e^{-iwT/2}-e^{iwT/2})=\frac{2\sin(wT/2)}{w}.
\end{eqnarray}

\subsection{hyperbolic tangent}
The Fourier transformation of hyperbolic tangent is given by
\begin{eqnarray}
\textit{F}_w[\tanh(t)]=-i\pi~{\rm csch}\bigg(\frac{\pi w}{2}\bigg),
\end{eqnarray}
from which
\begin{eqnarray}
\textit{F}_w[\tanh(at)]=\int dt \tanh(at) e^{-wt}\nonumber\\
=\frac{1}{a}\int d(at) \tanh(at) e^{-\frac{w}{a}at}=-i\frac{\pi}{a}~{\rm csch}\bigg(\frac{\pi w}{2a}\bigg).
\label{tanh}
\end{eqnarray}
Eq.~(\ref{signum}) is easily proved from Eq.~(\ref{tanh}) as below:
\begin{eqnarray}
\textit{F}_w[{\rm sgn}(t)]=\lim_{a\rightarrow \infty}\textit{F}_w[\tanh(at)]=\lim_{a\rightarrow \infty}\frac{\pi}{ai}~{\rm csch}\bigg(\frac{\pi w}{2a}\bigg)\nonumber\\
=\lim_{a\rightarrow \infty}\frac{\pi}{ai}\frac{2}{\exp(\frac{\pi w}{2a})-\exp(-\frac{\pi w}{2a})}=\frac{2}{iw}.~~~~~
\end{eqnarray}

\end{document}